\documentclass[aps, prx, twocolumn,longbibliography]{revtex4-1}
\usepackage[latin1]{inputenc} 
\usepackage[T1]{fontenc}

\usepackage{graphicx}
\usepackage{framed}
\usepackage{esvect}
\usepackage{amsmath,amssymb,color,bm}
\usepackage{mathrsfs}
\usepackage[dvipsnames]{xcolor}

\usepackage{amsmath,amssymb,color,bm}

\definecolor{cream}{RGB}{222,217,201}

\newcommand\tb[1]{\textbf{#1}}

\newcommand\mc[1]{\mathcal{#1}}

\newcommand\dd{\textnormal{d}}
\newcommand\beq{\begin{equation}}
\newcommand\eeq{\end{equation}}
\newcommand\beqa{\begin{eqnarray}}
\newcommand\eeqa{\end{eqnarray}}

\newcommand\ti[1]{\textit{#1}}
\newcommand\tn[1]{\textnormal{#1}}
\newcommand\ts[1]{\textcolor{black}{#1}}

\newcommand\im[1]{\textnormal{Im}\left[#1\right]}
\def\Ha{\mathcal{H}}

\def\ch{\tn{cosh}}
\def\sh{\tn{sinh}}

\def\coth{\tn{cotanh}}

\def\q{\tb{q}}

\def\x{\tb{r}}
\def\A{\mc{A}}
\newcommand{\hbarr}{\mathchar'26\mkern-9mu h}

\begin{document}

\title{Collective modes and quantum effects in two-dimensional nanofluidic channels}

\author{Baptiste Coquinot\textit{$^{a,b,c}$}}
\author{Maximilian Becker\textit{$^{d}$}}
\author{Roland R. Netz\textit{$^{d}$}}
\author{Lyd\'eric Bocquet\textit{$^{a}$}}
\author{Nikita Kavokine\textit{$^{b,c}$}}
\email{nikita.kavokine@mpip-mainz.mpg.de}

\affiliation{\textit{$^{a}$~Laboratoire de Physique de l'\'Ecole Normale Sup\'erieure, ENS, Universit\'e PSL, CNRS, Sorbonne Universit\'e, Universit\'e Paris Cit\'e, 24 rue Lhomond, 75005 Paris, France.}\\
\textit{$^{b}$~Department of Molecular Spectroscopy, Max Planck Institute for Polymer Research, Ackermannweg 10, 55128 Mainz, Germany.} \\
\textit{$^{c}$~Center for Computational Quantum Physics,
Flatiron Institute, 162 5$^{th}$ Avenue, New York, NY 10010, USA.}\\
\textit{$^{d}$~Fachbereich Physik, Freie Universit\"at Berlin, Arnimallee 14, 14195 Berlin, Germany.}
}

\date{\today}

\begin{abstract}
Nanoscale fluid transport is typically pictured in terms of atomic-scale dynamics, as is natural in the real-space framework of molecular simulations. An alternative Fourier-space picture, that involves the collective charge fluctuation modes of both the liquid and the confining wall, has recently been successful at predicting new nanofluidic phenomena such as quantum friction and near-field heat transfer, that rely on the coupling of those fluctuations. Here, we study the charge fluctuation modes of a two-dimensional (planar) nanofluidic channel. Introducing \emph{confined response functions} that generalize the notion of surface response function, we show that the channel walls exhibit coupled plasmon modes as soon as the confinement is comparable to the plasmon wavelength. Conversely, the water fluctuations remain remarkably bulk-like, with significant confinement effects arising only when the wall spacing is reduced to 7 \AA. We apply the confined response formalism to predict the dependence of the solid-water quantum friction and thermal boundary conductance on channel width for model channel wall materials. Our results provide a general framework for Coulomb interactions of fluctuating matter in nanoscale confinement. 
\end{abstract}

\maketitle


\section{Introduction}\label{Intro}
Fluids confined at the nanometer scale underly many technologically important processes \cite{Faucher2019,Bocquet2020}, including filtration, seawater desalination \cite{Phillip2011,Werber2016}, blue energy harvesting \cite{Logan2012,Siria2017} and electrochemical energy storage \cite{Chmiola2006}. Yet, they started to be fundamentally investigated not more than 20 years ago, and their initial theoretical description was largely inherited from macroscopic hydrodynamics, with generic walls imposing the same boundary conditions regardless of their material composition \cite{Kavokine2021}. The first nanofluidic effects emerged from the realization that, at the nanoscale, one may not neglect the wall's surface charge \cite{Schoch2005}, which results in coupled ion-fluid transport phenomena such as electro-osmosis and streaming currents \cite{Marbach2019}. There has been, however, accumulating evidence in recent years that surface charge is not a sufficient descriptor for the nanofluidic solid-liquid interface. From fluids near conducting surfaces \cite{Marin2020,Schlaich2022} to strongly interacting ions due to dielectric contrast \cite{Kavokine2019,Robin2021,Kavokine2022b}, several studies pointed to the need of describing the solid walls at the level of their electronic properties.

It may indeed be expected that, close enough to a solid wall, the Coulomb potentials produced by charged particles in a liquid are screened by the dielectric response of the wall material: this effect has been termed \emph{interaction confinement} \cite{Kavokine2022b}. Charged particles in a liquid are, first and foremost, ions: interaction confinement produces effective Coulomb interactions between ions in nanochannels that are modified compared to bulk Coulomb interactions, leading to a wealth of correlation effects \cite{Kavokine2019,Robin2021}. But a polar liquid such as water, even though electrically neutral after time-avergaging, has a molecular-level charge structure: water thus exhibits thermal charge fluctuations at terahertz frequencies and on a wide range of length scales \cite{Carlson2020} (termed \emph{hydrons} \cite{Coquinot2023}). The corresponding Coulomb fields are also subject to interaction confinement: they are dynamically screened by the thermal and quantum fluctuations of the electrons in the solid wall \cite{Volokitin2007,Kavokine2022}. This solid-liquid coupling has been shown to result in a "quantum" contribution to hydrodynamic friction, and in direct near-field energy transfer between the liquid and the solid's electrons \cite{Kavokine2022,yu2023}. These effects bridge the gap between fluid dynamics and condensed matter physics, opening the way to engineering nanoscale flows with the confining walls' electronic properties \cite{Coquinot2023,Lizee2023}. 

Fluctuation-induced effects in nanofluidics have so far been studied at the level of a single planar interface. The relevant many-body electrostatics were conveniently described in terms of surface response functions: surface analogues of the dielectric function that had been widely used, for instance, in the field of plasmonics \cite{Liebsch,Pitarke2007}. Here, we introduce \emph{confined response functions}, that generalize surface response functions to a 2D nanochannel geometry (Fig. 1a), providing a general tool for the treatment of Coulomb interactions in 2D confinement. As an illustration, we study the confined response of water, and of solid walls described as either graphene sheets or jellium slabs \cite{Lang1970,Griffin1976}: this allows us to predict the confinement dependence of solid-water quantum friction and thermal boundary conductance. 

This paper is organised as follows. 
In Sec.~\ref{correlation}, we introduce confined response functions and link them to eigenmodes of the Coulomb potential in the 2D nanochannel geometry.  
In Sec.~\ref{solid-water}, we compute the confined responses of specific media. We take the examples of a graphene sheet and a semi-infinite jellium for the solid; for the liquid, we study water in the framework of both force field and \emph{ab initio} molecular dynamics (MD) simulations.
In Sec.~\ref{quantum_friction}, we describe the influence of confinement on fluctuation-induced effects, specifically quantum friction and near-field heat transfer. To do so, we carry out the field-theory derivation of quantum friction directly in the confined geometry, which leads to natural emergence of the confined response functions. 
Finally, Sec.~\ref{conclusion} establishes our conclusions.

\paragraph*{Units and conventions.} We set the Boltzmann constant $k_{\rm B}=1$ (that is, we express the temperature in energy units), but otherwise use SI units throughout the text.  In real space, we use the cylindrical coordinates $\x=(\rho,z)$. The interfaces are at $z=0$ for a single-interface and at $z=\pm h/2$ for a confined channel.
 We use Fourier transforms for both $\rho$ and the time but never for the $z$-direction. We use the following convention for the $d$-dimensional Fourier transform:
\beqa 
\hat{F}(\tb{q},\omega)&=&\int \dd^d\tb{r}\dd t\, F(\tb{r},t)e^{-i\tb{q}\cdot\tb{r}+i\omega t},\nonumber\\
F(\tb{r},t)&=&\int \frac{\dd^d\tb{q}\dd\omega}{(2\pi)^{d+1}}\, \hat{F}(\tb{q},\omega)e^{i\tb{q}\cdot\tb{r}-i\omega t}.\nonumber
\eeqa

The charge densities are expressed in units of $e$ and the electrical potentials include an additional factor $e$. We denote $V(\tb{r})=e^2/(4\pi\epsilon_0 r)$ the Coulomb potential which becomes $V(\tb{q},z)=e^2/(2\epsilon_0 q)e^{-q|z|}$ in Fourier space.

\section{Electric response of interfacial systems}\label{correlation}

\begin{figure*}
    \centering
	\includegraphics[scale=0.8]{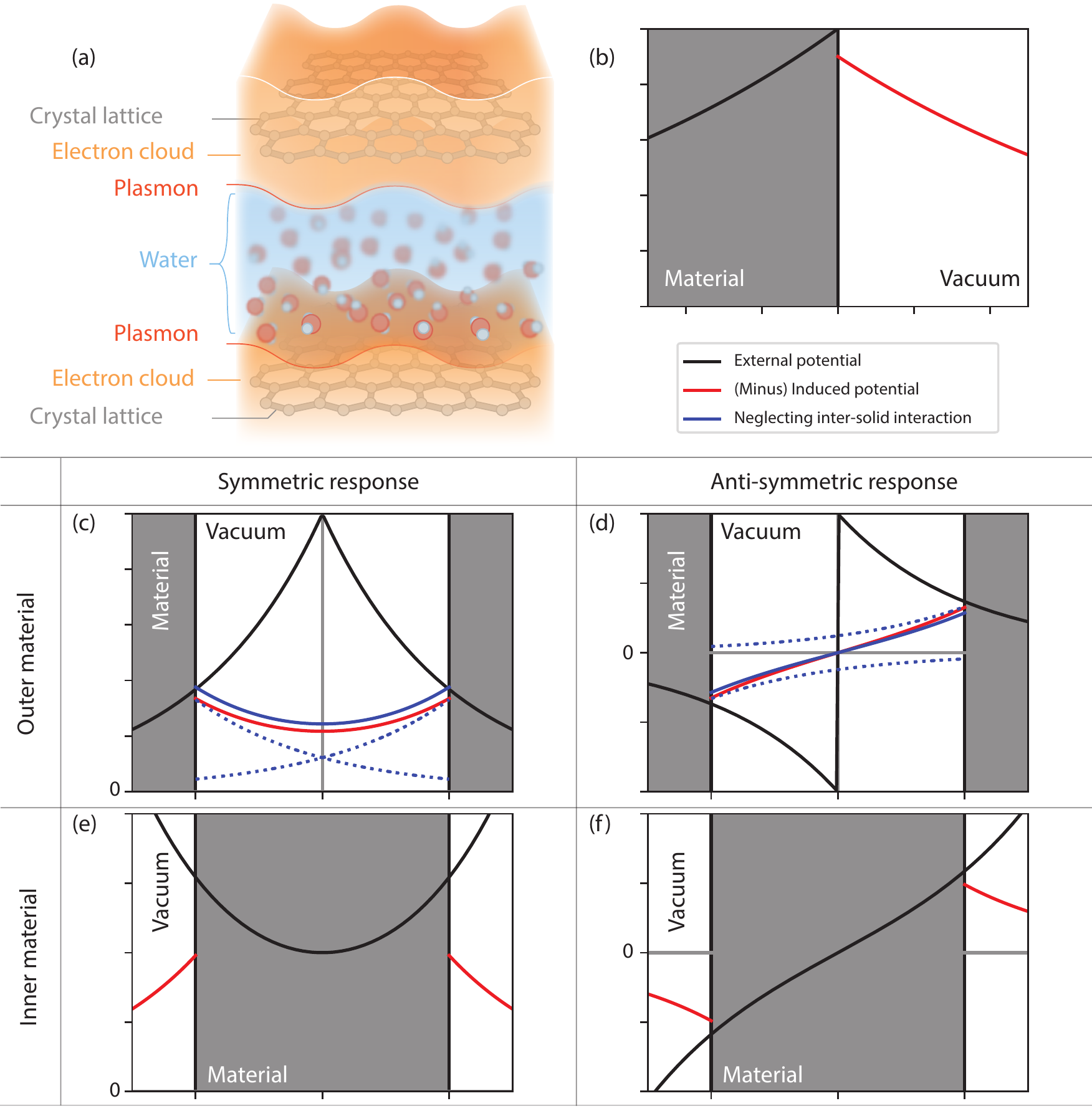}
	
    \caption{\emph{Electric response of interfacial systems.} a) Schematic of the confined geometry under consideration, emphasizing the role of the channel walls' electronic degrees of freedom. b-f): Electric response in different geometries. b) Single-interface case: the response function is the usual surface response function. c-f) In confined geometry there are four situations to distinguish. The responding medium can be either in the outer space (c\&d) or in the inner space (e\&f). The external potential can be either symmetric (c\&e) or antisymmetric (d\&f). We observe that the inter-solid interactions bring corrections by lowering the induced potential in the symmetric case (c) and increasing the induced potential in the antisymmetric case (d).}
    \label{fig1}
\end{figure*}

 \begin{table*}
 \centering
 \renewcommand{\arraystretch}{2}
\begin{tabular}{ l @{\hspace{0.5cm}} l @{\hspace{0.5cm}} @{\hspace{0.5cm}}c @{\hspace{0.5cm}} @{\hspace{0.5cm}}c@{\hspace{0.5cm}} }
    \hline  
    \tb{Model} & \tb{Geometry} & \multicolumn{2}{c}{\tb{Weight function}}  \\
    \hline  
    \tb{Single interface} & \underline{Half-space}  & \multicolumn{2}{c}{$F^0(q,z)=e^{-q|z|}$}   \\ \hline 
   \tb{Confined} &  & \underline{Symmetric} &   \underline{Antisymmetric}  \\
     & \underline{Inside} &$F^{\rm s}_{\rm i}(q,z)=\sqrt{2}\,\ch(qz)e^{-qh/2}$ & $F^{\rm a}_{\rm i}(q,z)=\sqrt{2}\,\sh(qz)e^{-qh/2}$\\
     & \underline{Outside} &  $F^{\rm s}_{\rm o}(q,z)=\frac{1}{\sqrt{2}}e^{-q(|z|-h/2)}$ & $F^{\rm a}_{\rm o}(q,z)=\frac{\tn{sign}(z)}{\sqrt{2}}e^{-q(|z|-h/2)}$\\
     \hline 
     \end{tabular}
\caption{Weight functions used in the definitions of the surface and confined response functions (Eq. \eqref{def_g0} and \eqref{IRF}).}\label{able}
\end{table*}

\subsection{Single-interface: surface response function}

We first briefly recall the widely-used concept of surface response function \cite{Liebsch,Pitarke2007}. Consider a semi-infinite medium occupying the half-space ($z < 0$). Given the electrostatic potential $\phi_{\rm ext}$ applied by an external source (an appropriate charge distribution) inside the medium, we wish to determine the potential $\phi_{\rm ind}$ induced by the medium in the half-space $z>0$. The potential $\phi_{\rm ext}$ must solve the Laplace equation for $z<0$. The physically-meaningful (non-diverging) solutions are given by the evanescent plane waves 
\beqa  \phi_{\rm ext}(\q,z,\omega)=\phi_{\rm ext}(\q,z=0,\omega)F^0(q,z), \nonumber\\   F^0(q,z)=e^{-q|z|}, \eeqa
where we have introduced the surface weight function $F^0$: this seemingly cumbersome notation will be useful upon generalization to a confined geometry. Assuming that the medium has a linear charge density response function $\chi$, the induced potential is given by (see Fig. 1b)
\beq  \phi_{\rm ind}(\q,z,\omega)=-g^0(\tb{q},\omega)\phi_{\rm ext}(\q,z=0,\omega)F^0(q,z)  \eeq
where we have introduced the \emph{surface response function} 
\beq \label{def_g0}
g^0(\tb{q},\omega)=-\frac{e^2}{2\epsilon_0 q}\int\dd z\dd z'\, F^0(q,z)\chi(\tb{q},z,z',\omega)F^0(q,z').
\eeq
It is worth noting that $g^0(\tb{q},\omega)$ is a scalar. Since we are considering a linearly responding medium, the response to an evanescent plane wave at $(\tb{q},\omega)$ is an evanescent plane wave at $(\tb{q},\omega)$, so that the induced potential has the same $z$-dependence as the external potential, given by the weight function $F^0$. In other words, the evanescent plane waves form an eigenbasis for the surface response. 

Let us illustrate the role of the surface response function with a simple example. We consider a static point charge $e$ at at distance $z_0$ from the interface. This charge produces an "external" potential 
\beq  \phi_{\rm ext}(\rho,z)=\int \frac{\dd\q}{(2\pi)^2}\frac{e^2}{2 q}F^0(q,z-z_0)e^{i\rho\cdot\q}  \eeq 
where we have introduced a Fourier decomposition into evanescent plane waves. 
Accounting for the response of the medium, the total potential is 
\beq  \phi(\q,z)=\frac{e^2}{2 q}\left[F^0(q,z-z_0)-g^0(\tb{q})F^0(q,z_0)F^0(q,z)\right]  \eeq
The surface response function gives the contribution of the medium's polarization to the total potential.

\subsection{Double interface: confined response functions}

We now generalize this approach to a two-dimensional nanochannel geometry. We consider two interfaces at $z=\pm h/2$ that define the channel, with the outside medium ($|z| > h/2$) being distinct from the inside medium ($|z| < h/2$). Later, we will specify that we consider water inside the channel, but we remain general at this point. 

Contrary to the previous case, there is no longer a symmetry between the two media, and we need therefore to distinguish the responses of the inner and outer medium. All the applied and induced potentials still satisfy the Laplace equation, but, in both media, the subspace of harmonic functions with wavevector $\tb{q}$ and frequency $\omega$ is now of dimension 2: the response function at a given $(\tb{q},\omega)$ is then in principle given by a $2\times 2$ matrix. We shall, however, express the potentials in a basis of even (symmetric) and odd (antisymmetric) harmonic functions, where the matrix turns out to be diagonal. We call these basis functions \ti{confined weight functions} and denote them $F^{\rm s/a}_{\rm i/o}$, where s/a stands for symmetric/antisymmetric function and i/o for inner/outer medium (see Table 1). The amplitude of the basis functions is in principle arbitrary: it is chosen so that the confined response functions defined in the following reduce to the conventional surface response functions in the non-confined case. 

Let us start with the response of the inner medium. We consider a generic external potential $\phi_{\rm ext}(\tb{q},z,\omega)=\phi^s(\q,\omega) F^{\rm s}_{\rm i}(q,z)+\phi^a(\q,\omega) F^{\rm a}_{\rm i}(q,z)$ applied on the inner medium of charge susceptibility $\chi_{\rm i}$. The induced potential is then 
\beqa \phi_{\rm ind}(\tb{q}, z)=\frac{e^2}{2\epsilon_0 q}\int_{-h/2}^{h/2} \dd z'\dd z''\, e^{-q|z-z'|}\ldots \nonumber\\\ldots\chi_{\rm i}(\tb{q},z',z'')\phi_{\rm ext}(\tb{q},z'').\eeqa
In the outer space $|z|>h/2$, taking advantage of the definition of the confined weight functions (see Table 1), this reduces to 
\beq  \phi_{\rm ind}(\tb{q}, z)=-g^{\rm s}_{\rm i}\phi^s F^{\rm s}_{\rm o}(q,z)-g^{\rm a}_{\rm i}\phi^a F^{\rm a}_{\rm o}(q,z) \eeq
where we have defined a generic \emph{confined response function}:
\beqa\label{IRF}
g_{\rm m}^{\rm c}(\tb{q},\omega)=-\frac{e^2}{2\epsilon_0 q}&&\int\dd z\dd z'\, F^{\rm c}_{\rm m}(q,z)\ldots \nonumber\\&&\ldots\chi_{\rm m}(\tb{q},z,z',\omega)F^{\rm c}_{\rm m}(q,z').
\eeqa
where the interaction takes place over the domain of definition of the weight function $F^{\rm c}_{\rm m}$; $\rm m = i, o$ and $\rm c = s, a$. 
To summarise, the inner medium responds to a perturbation of the form $F^{\rm s/a}_{\rm i}$ by a potential of the form $F^{\rm s/a}_{\rm o}$ in the outer medium with an amplitude $-g^{\rm s/a}_{\rm i}$ (see Fig. \ref{fig1}c-d). 

We proceed similarly for the response of the outer medium (with charge susceptibility $\chi_{\rm o}$). It responds to an external potential $\phi_{\rm ext}(\tb{q},z,\omega)=\phi^s(\q,\omega) F^{\rm s}_{\rm o}(q,z)+\phi^a(\q,\omega) F^{\rm a}_{\rm o}(q,z)$ with an induced potential in the inner space $|z|<h/2$:
\beq  \phi_{\rm ind}=-g^{\rm s}_{\rm o}\phi^s F^{\rm s}_{\rm i}(z)-g^{\rm a}_{\rm o}\phi^a F^{\rm a}_{\rm i}(z). 
\label{phiindo}
\eeq
The two components of the outer medium response are shown in Fig. \ref{fig1}e-f. 

To summarize, we have generalized surface response functions to a 2D nanochannel geometry by identifying the new eigenmodes of the Coulomb potential, given by the confined weight functions. The response to a symmetric (antisymmetric) weight function is a symmetric (antisymmetric) weight function with amplitude given by the confined symmetric (antisymmetric) response function, in the same way that the response to an evanescent wave is an evanescent wave in the single interface case. 

\subsection{Fluctuation-dissipation theorem and physical interpretation}\label{FDT}

To obtain a physical interpretation of the confined response functions, it is useful to relate them to equilibrium charge density fluctuations using the fluctuation-dissipation theorem (FDT). For the charge susceptibility $\chi$, the FDT reads \cite{Kavokine2022}
\beq S_{\rm m}(\x, \x',\omega)=f(\omega)\im{\chi_{\rm m}(\x,\x',\omega)}\eeq
where $f(\omega)=2T/\omega$ for classical dynamics and $f(\omega)=\hbarr\,\coth(\hbarr\omega/(2T))$ for quantum dynamics. The structure factor $S$ is defined as 
\beq
S_{\rm m}(\x,\x',t)=\langle n_{\rm m}(\x,t)n_{\rm m}(\x',0)\rangle, 
\eeq
where $n_{\rm m}$ is the charge density of the medium.
Adapted to the confined response function defined in Eq. \eqref{IRF}, the FDT becomes 
\beq S^{\rm c}_{\rm m}(\q,\omega)=f(\omega)\im{g_{\rm m}^{\rm c}(\tb{q},\omega)}\eeq
where $S^{\rm c}$ is a confined structure factor defined as 
\beqa 
S^{\rm c}_{\rm m}(\q,t)=\frac{1}{\A}\int\dd\x\dd \x'\, \langle n_{\rm m}(\x,t)n_{\rm m}(\x',0)\rangle\ldots \nonumber\\\ldots e^{-i\q\cdot(\rho-\rho')}F^{\rm c}_{\rm m}(q,z)F^{\rm c}_{\rm m}(q,z')
\label{FDTc}
\eeqa
where $\A$ is the area of the interface.  

In Eq.~\eqref{FDTc}, if the weight function $F_{\rm m}^{\rm c}$ is symmetric, the charge density $n_{\rm m}(z)$ can be replaced with $(n_{\rm m}(z)+n_{\rm m}(-z))/2$: the structure factor only counts the symmetric charge fluctuations. Similarly, if the weight function is antisymmetric, the charge density $n_{\rm m}(z)$ can be replaced with $(n_{\rm m}(z)-n_{\rm m}(-z))/2$: the structure factor only counts the antisymmetric charge fluctuations. Thus, for the channel walls, the symmetric (antisymmetric) response function accounts for in-phase (out-of-phase) coupled modes. For the inner medium, the symmetric (antisymmetric) response function accounts for monopolar (dipolar) charge fluctuations. 

\subsection{Confined response function of the outer medium}\label{outer}

We now provide an expression of the outer medium confined response function in terms of the usual surface response functions. 
We distinguish the contributions of the two solid walls to the potential induced in the inside medium: we denote them $\phi_{\rm ind}^{\rm T}$ and $\phi_{\rm ind}^{\rm B}$ for the "top" and "bottom" solids, respectively. We need to account for the solid-solid interactions: each solid responds to the external potential and to the potential induced by the other solid. Let us consider an external potential 
$\phi_{\rm ext}(\tb{q},z,\omega)=\phi_{\rm o}(\q,\omega) F^{\rm c}_{\rm o}(q,z)$ where the weight function $F^{\rm c}_{\rm o}$ is either $ F^{\rm s}_{\rm o}$ or $ F^{\rm a}_{\rm o}$. 
The top solid induces a potential
\beqa \phi_{\rm ind}^{\rm T}(z)&=&-g^0_{\rm o}F^0(z-h/2)\ldots \nonumber\\&&\ldots\left[\phi_{\rm o} F^{\rm c}_{\rm o}(z=h/2)+\phi_{\rm ind}^{\rm B}(z=h/2)\right]\eeqa
for $z<h/2$, while the bottom solid induces a potential 
\beqa \phi_{\rm ind}^{\rm B}(z)&=&-g^0_{\rm o}F^0(z+h/2)\ldots \nonumber\\&&\ldots\left[\phi_{\rm o} F^{\rm c}_{\rm o}(z\! =\! -\! h/2)+\phi_{\rm ind}^{\rm T}(z\! =\! -\! h/2)\right]\eeqa
for $z>-h/2$. Combining these two equations, we obtain the total induced potential as 
\beq \phi_{\rm ind}(z)=\phi_{\rm ind}^{\rm T}(z)+\phi_{\rm ind}^{\rm B}(z)=-g^{\rm c}_{\rm o}F^{\rm c}_{\rm i}(z)\phi^0\eeq
in the inner space $|z|<h/2$. Comparing to Eq.~\eqref{phiindo}, we deduce 
\beq \label{gco}
g^{\rm s}_{\rm o}=\frac{g^0_{\rm o}}{1+g^0_{\rm o}e^{-qh}}, \quad \quad  g^{\rm a}_{\rm o}=\frac{g^0_{\rm o}}{1-g^0_{\rm o}e^{-qh}}.
\eeq
We find that the confined response functions reduce to surface response functions at wavelengths $1/q \ll h$: the deviation from the surface response function originates from the Coulomb interaction between the two walls. As expected on physical grounds, the inter-wall interaction reduces the in-phase (symmetric) response and enhances the out-of-phase (antisymmetric) response. However, the induced potential does not exceed the applied potential in confinement if it does not for a single interface: there is no confinement-induced overscreening (see \ts{ESI.1}). 

\subsection{Interaction confinement}

A first consequence of the wall electric response in a 2D channel is a modification of the effective Coulomb interactions between charged particles inside the channel: this effect has been termed \emph{interaction confinement} \cite{Kavokine2022b}. Two confined charges (for example, ions in water) interact not only directly, but also indirectly, through the polarization charges induced in the channel walls. Coulomb interactions near polarizable walls have been computed in various geometries and within different models for the wall polarizability.  

In ref. \cite{Kavokine2022b}, interaction confinement in a 2D channel geometry was addressed in the framework of surface response functions, which allowed for evaluation of effective Coulomb interactions between two ions in the channel mid-plane, for various models of the channel wall material. For example, with walls described by a Thomas-Fermi model,  the ion-ion interaction is reinforced with respect to bulk water at small distances, and exponentially screened at large distances. 

Our confined response formalism generalizes the results of ref. \cite{Kavokine2022b}, allowing for the evaluation of the effective Coulomb interactions between arbitrary charge distributions. Any charge distribution can indeed be decomposed into an even and an odd part. The wall response to the even (odd) part is then given by the symmetric (antisymmetric) response function. In \ts{ESI.2}, we provide expressions in terms of the confined response functions for the effective Coulomb potential produced by prototypical even (point charge) and odd (dipole) charge distributions. 

\section{Confined response of model media}\label{solid-water}

\begin{figure*}
    \centering
	\includegraphics{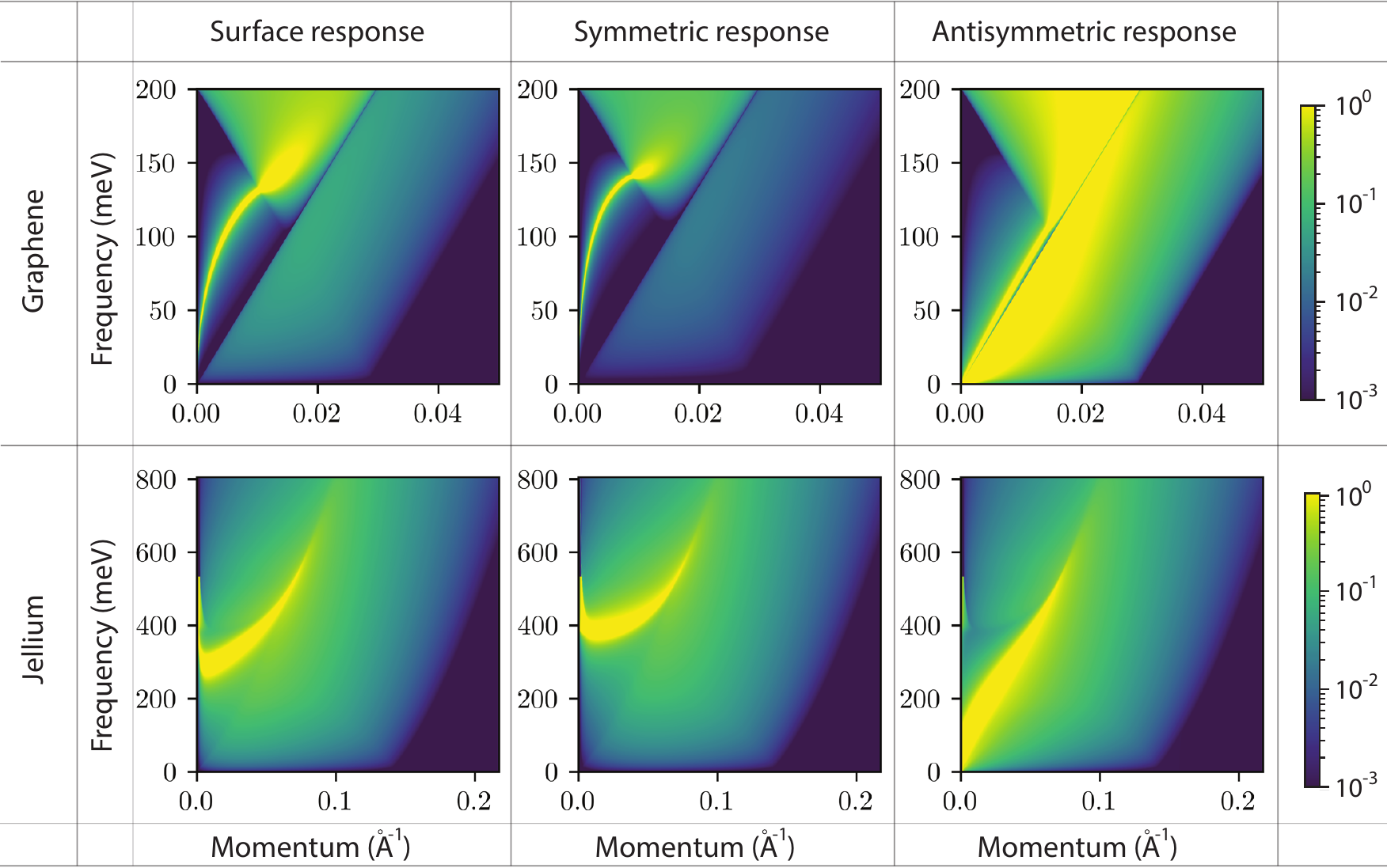}
	
    \caption{\emph{Surface and confined response functions for different solids.} a-f) Interfacial response functions Im$[g(q,\omega)]$ of the solid as a function of momentum $q$ and frequency $\omega$. The first row (a-c) is for graphene with Fermi level $\mu~=~100$~meV.
    The second row (d-f) is for a jellium model with an effective mass $m~=~0.1~m_{\rm e}$ and a Fermi level $\mu~=~180$~meV (electron density parameter $r_s~=~5$), corresponding to a doped semi-conductor.  
    The first column (a\&d) corresponds to the surface response functions. 
    The second column (b\&e) corresponds to the symmetric confined response functions with a confinement of 7~\AA. 
    The third column (c\&f) corresponds to the anti-symmetric confined response functions with a confinement of 7~\AA. }
    \label{fig2}
\end{figure*}

\begin{figure*}
    \centering
	\includegraphics[scale=1]{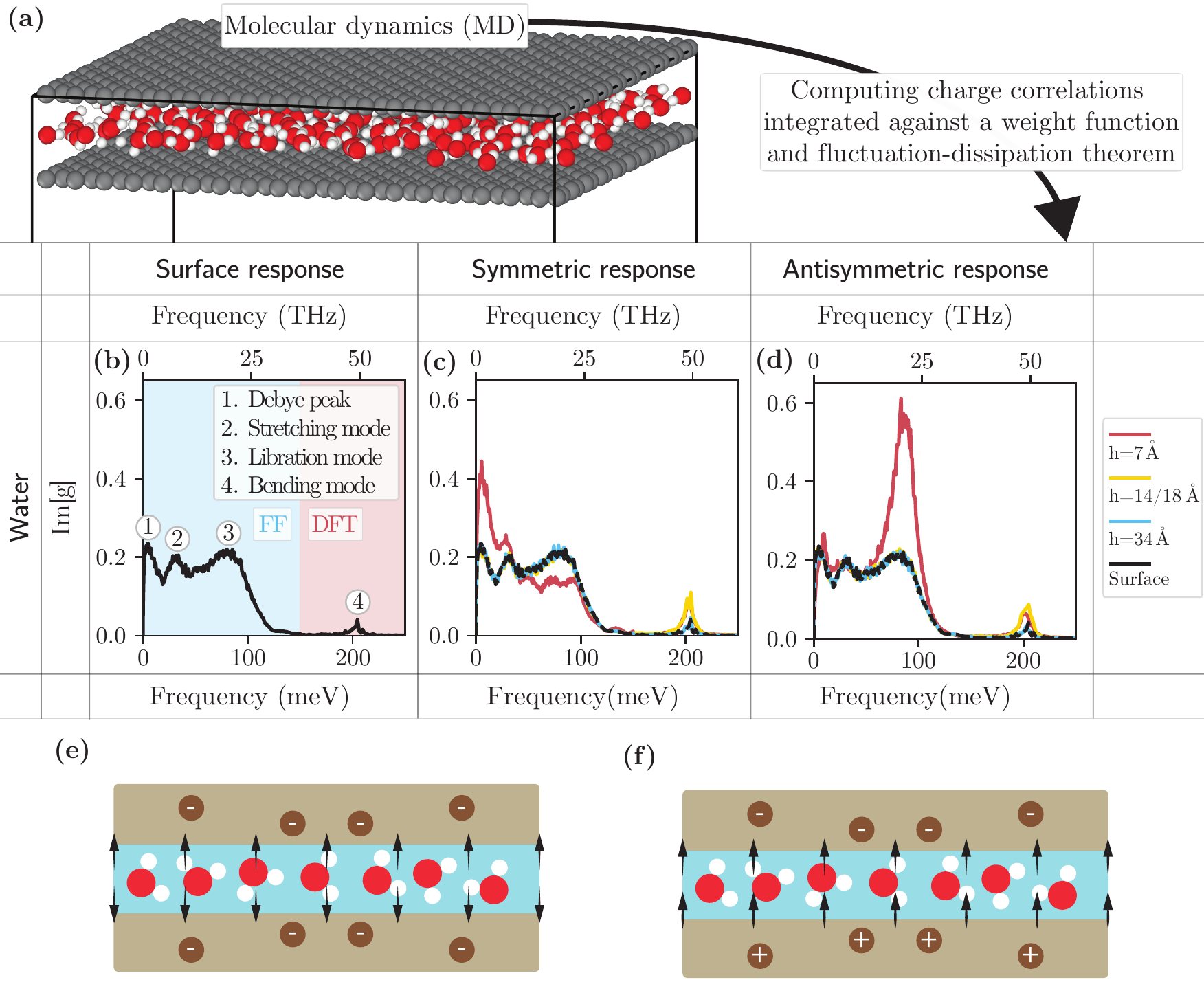}
	
    \caption{\emph{Water response functions from simulation.}    a) Snapshot of a FF molecular dynamics simulation with 7~\AA~confinement b-d) Response functions Im$[g(q,\omega)]$ of water as a function of the frequency $\omega$ obtained from the simulations, at fixed wavevector $q~=~0.67~\tn{\AA}^{-1}$ for FF and $q~=~1~\tn{\AA}^{-1}$ for DFT. Different confinements are used: 7~\AA,  14~\AA~(only FF), 18~\AA~(only DFT), 34~\AA~and 60~\AA~(only FF). b) Surface response function computed from simulations at weak confinement (FF at $h=$60~\AA~and DFT at $h=$34~\AA). c) Symmetric response function for different confinements. The dashed black line is the non-confined surface response function. d) Antisymmetric response function for different confinements. The dashed black line is the non-confined surface response function. e) Schematic of a channel containing a single water monolayer on which a symmetric potential is applied. f) Same as (e) with application of an antisymmetric potential.}
    \label{fig3}
\end{figure*}

We now specialise to a 2D nanofluidic channel of height $h$ filled with water, and we investigate the effect of confinement on the electric response of water and of model channel wall materials. 
\subsection{Outer medium: coupled plasmon modes}

We consider two models for the channel wall material: a graphene monolayer and a semi-infinite jellium. The graphene surface response function is computed as detailed in ref. \cite{Kavokine2022}, starting from the charge susceptibility in the Dirac cone and zero-temperature approximations, at a chemical potential $\mu~=~100~\rm meV$. The semi-infinite jellium is treated in the specular reflection approximation, as detailed also in ref. \cite{Kavokine2022}. In the jellium model, electrons are free in a uniform positive background, and are completely characterised by their chemical potential $\mu$ and effective mass $m$. We use $\mu~=~180~\rm meV$ and $m~=~0.1~\rm m_{\rm e}$ as a model for a doped semi-conductor: this corresponds to an electron density parameter $r_{\rm s}~=~5$. 

For both systems, the surface response functions feature a sharply-defined surface plasmon mode and a broad particle-hole continuum (Fig.~\ref{fig2}). The effect of confinement is most clearly visible on the plasmon mode: its energy is increases in the confined symmetric response and decreased in the confined asymmetric response. This is consistent with the physical interpretation outlined above: as the two solid walls face each other, in-phase (out-of-phase) charge density oscillations have an increased (decreased) energy cost due to the Coulomb interactions between the two solids. These interactions are significant only for charge fluctuations whose wavelength is longer than the confinement width $h$, so that the confined response functions differ from the surface response function only at small enough momenta $q$ (see Eq.~\eqref{gco}). 

We anticipate that the formation of coupled plasmon modes between the walls of a 2D nanofluidic channel will affect transport inside the channel, and particularly fluctuation-induced effects (see Sec. \ref{quantum_friction}).

\subsection{Inner medium: confined water spectra from simulations}\label{water}

We now turn to the confined response functions of water. Confinement may impact water charge fluctuations in two ways. First, the interaction between the two interfaces of the water slab is expected to result in a difference between the symmetric and antisymmetric responses. Second, the confinement-induced modifications of the water structure may intrinsically affect its charge fluctuations. 

We determine the water response functions in the framework of molecular dynamics (MD) simulations. We carry out both force-field (FF) and \emph{ab intio} density functional theory-based (DFT) simulations of water confined between two frozen graphene sheets, for various separations $h$ between the graphene sheets. From the simulation trajectories, we compute the charge structure factor of water, integrated along $z$ after multiplication by the weight functions summarized in Table 1, and determine the corresponding response functions through the fluctuation-dissipation theorem (Sec. \ref{FDT}). DFT simulations are required to capture intra-molecular modes: we use them to obtain the spectra at frequencies above 150 meV. At lower frequencies, we use the FF simulations to capture the contribution of inter-molecular modes, inaccessible with the short simulation times of DFT. Details of the numerical parameters and procedures are given in \ts{ESI.3}. 

The results are presented in Fig.~\ref{fig3}. Fig.~\ref{fig3}b shows the water surface response function (obtained from the simulation at weakest confinement) at a fixed wavevector $q_0 = 0.67~\tn{\AA}^{-1}$ and in the frequency range $0-200~\rm meV$. In this range, the water charge response can be decomposed into four modes: in order of increasing frequency, the Debye mode, the hydrogen-bond stretching mode, the libration mode and the OH-bond bending mode \cite{Bonthuis2012,Carlson2020}. We note that the OH-stretch mode (at around 450 meV) falls outside the studied frequency range. The confinement dependence of these modes could in principle be analysed in terms of their molecular origin; this is, however, beyond the scope of this article, and we restrict ourselves to a phenomenological description. 

Overall, the water response functions are remarkably robust to confinement. Down to $h = 1.4~\rm nm$, the lower-frequency intermolecular modes remain unaffected. We observe, however, a slight red shift, and an increased oscillator strength for the bending mode. A significant effect on the intermolecular modes is visible only at 7 \AA~confinement. In the symmetric response, the Debye and hydrogen-bond-stretch modes are amplified, while the libration mode is suppressed; in the antisymmetric response, the libration peak is strongly amplified while the other modes are unaffected. As illustrated in Figs. \ref{fig3}e-f, in the symmetric case, the perpendicular component of the applied electric field changes direction across the channel, while it maintains a constant sign in the antisymmetric case. This likely indicates that the libration mode is mostly excited by the electric field perpendicular to the interface, while the lower frequency modes are excited by the parallel component. 

\begin{figure}
     \centering
     \includegraphics[scale=1]{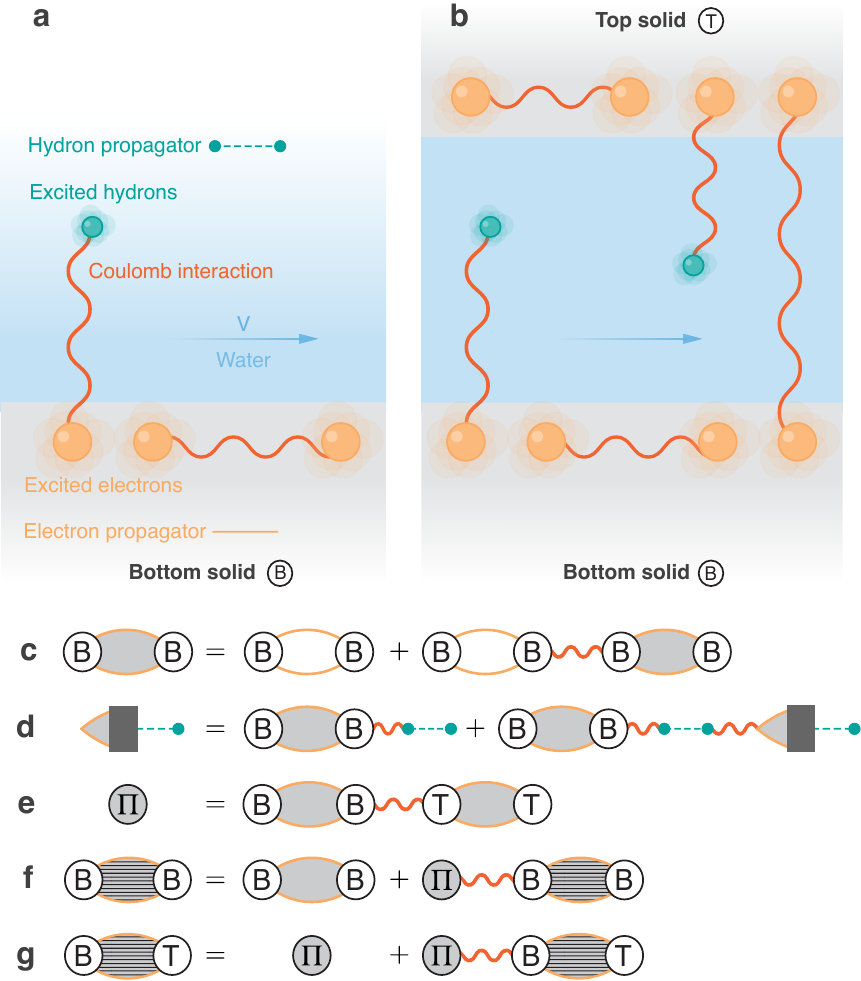}

     \caption{\emph{Fluctuation-induced effects.} 
     \tb{(a)} Schematic of the interactions involved in single-interface quantum friction. Friction results from the Coulomb coupling of liquid fluctuations (hydrons) and electronic fluctuations. 
     \tb{(b)} Schematic of the interactions involved in confined quantum friction. Compared to the non-confined case, there is an additional Coulomb interaction between the charge fluctuations in the two solid walls.  
      \tb{(c)} Dyson equation for the renormalization of the bottom wall's charge density response function (polarization "bubble") by the intra-wall Coulomb interactions. The bare bubble is empty and the renormalized bubble is filled with gray. 
          \tb{(d)} Dyson equation for the electron-water susceptibility $\chi_{\rm ew}$. 
    \tb{(e)} Definition of the exchange term $\Pi$ (see text). 
    \tb{(f)} Dyson equation for the renormalization of the bottom wall's charge density response function by the inter-wall Coulomb interactions. The stripes indicate a fully-renormalized bubble. 
   \tb{(g)} Dyson equation for the inter-wall charge density response function, which vanishes in the absence of inter-wall Coulomb interactions ($\Pi = 0$). }
     \label{fig4}
 \end{figure}

\section{Confined quantum friction and heat transfer}\label{quantum_friction}
 
In this section, we discuss the effect of 2D confinement on fluctuation-induced interfacial effects -- solid-liquid quantum friction and near-field radiative heat transfer -- making use of the confined response function formalism developed above.

\subsection{Single-interface quantum friction }

We start by briefly summarising the physics of quantum hydrodynamic friction. The classical friction between a liquid and a solid is usually determined by the solid's surface roughness \cite{Barrat1999,Huang2008}. However, it was recently shown that the classical contribution is supplemented by a fluctuation-induced or "quantum" contribution, due to the coupling of water charged fluctuations (termed \emph{hydrons}) to electronic excitations within the solid \cite{Kavokine2022}. When undergoing quantum friction, a liquid transfers momentum directly to the solid's electrons. Similarly, a liquid may transfer energy directly to the solid's electrons: this is near-field radiative heat transfer \cite{Biehs2021,yu2023}. 

In the case of a single solid-liquid interface, if the liquid is flowing at velocity $\tb{v}$, it is subject to a quantum friction force $\tb{F} = - \lambda \mathcal{A} \tb{v}$ with the quantum friction coefficient $\lambda$ given by \cite{Kavokine2022} 
\beq\label{lambda_force}
\lambda =\frac{\hbarr^2}{8\pi^2T}\int_0^\infty\dd\omega\dd q\, \frac{q^3}{\sh^{2}\left(\frac{\hbarr\omega}{2T}\right)} \Delta \gamma[F^0].
\eeq
Anticipating the generalization to the confined case, we have introduced the notation 
\beq\label{dgamma}
\Delta \gamma[F]=\frac{\im{g_{\rm e}(q,\omega)}\im{g_{\rm w}(q,\omega)}}{|1-g_{\rm e}(q,\omega)g_{\rm w}(q,\omega)|^2},
\eeq
where the $g$'s ($e$ corresponds to the solid, $w$ to the liquid) are generalized response functions computed with the weight function $F$ (see Table 1). Similarly, if there is a temperature difference $\Delta T$ between the solid and the liquid, the solid-liquid heat flux is given by $\mathcal{Q} = \kappa \mathcal{A} \Delta T$, where the thermal boundary conductance $\kappa$ reads \cite{yu2023}
\beq
\kappa =\frac{\hbarr^2}{4\pi^2T^2}\int_0^\infty\dd\omega\dd q\, \frac{q\omega^2}{\sh^{2}\left(\frac{\hbarr\omega}{2T}\right)} \Delta \gamma[F^0].
\eeq

\subsection{Confined quantum friction }

We now generalize the above results to the 2D confined geometry presented in Fig.~\ref{fig4}b. We wish to compute the total quantum friction force applied by the solid walls on the flowing liquid, and the total heat transfer rate between the liquid and the two solid walls. The liquid and the solid's electrons are described by their fluctuating charge densities $n_{\rm w}$ and $n_{\rm e}$, which have a Coulomb interaction of the form 
\beq
\Ha_{\rm h-e}(t)=\int\dd\tb{r}_{\rm w}\dd\tb{r}_{\rm e}\, n_{\rm w}(\tb{r}_{\rm w},t)V(\tb{r}_{\rm w}-\tb{r}_{\rm e})n_{\rm e}(\tb{r}_{\rm e},t).
\eeq
The dynamics of the systems are governed by the interaction Hamiltonian that also comprises the electron-electron Coulomb interactions: $\Ha_{\rm int}=\Ha_{\rm e-e}+\Ha_{\rm h-e}$. 

The friction force and heat transfer rate are given by
\beq 
\langle \tb{F}\rangle =-\int\dd\tb{r}_{\rm w}\dd\tb{r}_{\rm e}\, \nabla V(\tb{r}_{\rm w}-\tb{r}_{\rm e}) \langle n_{\rm w}(\tb{r}_{\rm w}-\tb{v}t,t)n_{\rm e}(\tb{r}_{\rm e},t)\rangle,
\label{Fdef}
\eeq
\beq 
\langle \mc{Q}\rangle =\int\dd\tb{r}_{\rm w}\dd\tb{r}_{\rm e}\,  V(\tb{r}_{\rm w}-\tb{r}_{\rm e})\partial_t \langle n_{\rm w}(\tb{r}_{\rm w},t)n_{\rm e}(\tb{r}_{\rm e},t)\rangle,
\label{Qdef}
\eeq
where the integration over $\tb{r}_{\rm w}$ (resp. $\tb{r}_{\rm e}$) runs over the space occupied by the liquid (resp. solid). The correlation functions appearing in Eqs. \eqref{Fdef} and \eqref{Qdef} may be computed in perturbation theory with respect to $\Ha_{\rm int}$, as has been detailed in ref. \cite{Kavokine2022}. Since the system is subject either to liquid flow or to a temperature gradient, the perturbative expansion needs to be carried out in the non-equilibrium Keldysh formalism \cite{Rammer2007}. Ultimately, both the friction force and the heat transfer rate can be obtained in terms of the solid-liquid charge density correlation function $\chi_{\rm ew}$, which, upon resummation of the perturbation series, is found to satisfy the following Dyson equation: 
  \beq
  \chi_{\rm ew}=\chi_{\rm e}\ast\chi_{\rm w}+\chi_{\rm e}\ast\chi_{\rm w}\ast\chi_{\rm ew}. \label{Dyson}
  \eeq
Here $\ast$ stands for convolution in space and time, multiplication by the Coulomb potential, and contraction of the Keldysh indices. A Feynman diagram representation of Eq.~\eqref{Dyson} is given in Fig.~\ref{fig4}d. The Keldysh indices carried by the $\chi$'s are not important for the geometrical discussion that follows. We will therefore not write them out explicitly and we refer the reader to ref. \cite{Kavokine2022} for further details: once the space-time convolutions have been dealt with, the computation is completely analogous to ref. \cite{Kavokine2022}. 

In the single interface case, upon Fourier transformation in time and in space parallel to the interface, Eq.~\eqref{Dyson} immediately becomes a scalar equation for surface response functions. In the channel geometry, we need to introduce \emph{confined response functions}, first as $2\times 2$ matrices in the indices $\nu, \xi =$ T,B:
	\beqa
(g_{\rm ab})^{\nu\xi}(\tb{q},\omega) &=& -\frac{ e^2}{2\epsilon_0 q}\int_{\mathfrak{Z}_{\rm a}^\nu}\dd z_{\rm a}\int_{\mathfrak{Z}_{\rm b}^\xi}\dd z_{\rm b}\, \chi_{\rm ab}(z_{\rm a},z_{\rm b},\tb{q},\omega)\ldots\nonumber\\
&&\qquad \ldots e^{q[\epsilon^{\rm a}(\sigma^\nu z_{\rm a}-h/2)+\epsilon^{\rm b}(\sigma^\xi z_{\rm b}-h/2])}\eeqa
where a,b=e,w, $\epsilon^{\rm w}=+$, $\epsilon^{\rm e}=-$, $\sigma^{\rm T}=+$ and $\sigma^{\rm B}=-$. The space has been divided into three regions: the central region $\mathfrak{Z}_{\rm w}$, and the bottom (B) and top (T) wall regions: $\mathfrak{Z}_{\rm e}=\mathfrak{Z}_{\rm e}^{\rm B}\cup\mathfrak{Z}_{\rm e}^{\rm T}$. In the convolution over $z$ in Eq.~\eqref{Dyson}, summing over the top and bottom solids corresponds to summing over the indices $\rm B, T$. Thus, in terms of the confined response functions, Eq.~\eqref{Dyson} becomes 
  \beq
g_{\rm ew}=- g_{\rm e}\cdot g_{\rm w}+g_{\rm e}\cdot g_{\rm w}\cdot g_{\rm ew}, \label{Dyson_gab}
  \eeq
where the dot represents the matrix product, and $g_{\rm a} \equiv g_{\rm aa}$. Neglecting the off-diagonal terms in Eq.~\eqref{Dyson_gab}, we would recover two copies of the single interface Dyson equation, one for the top and one for the bottom interface. The off-diagonal terms represent cross-talk between the walls (for example, the top solid wall responding to a water fluctuation near the bottom wall), which is expected to vanish for weak confinement. 

We assume in the following that the top and bottom wall materials are the same, so that the system is symmetric under the mirror transformation $z\rightarrow -z$.  As a consequence, $g^{\rm TT}=g^{\rm BB}$ and $g^{\rm TB}=g^{\rm BT}$. Therefore, all the confined response matrices can be diagonalised in the form
\beq PgP^{-1}=\begin{pmatrix}
g^{\rm TT}+g^{\rm TB}  & 0\\
0  & g^{\rm TT}-g^{\rm TB}
\end{pmatrix}=
\begin{pmatrix}
g^{\rm s}  & 0\\
0  & g^{\rm a}
\end{pmatrix}\eeq 
where we have used 
\beq P=\frac{1}{\sqrt{2}}\begin{pmatrix}
1  & 1\\
1  & -1
\end{pmatrix},
\eeq 
which satisfies $P^{-1}=P.$ We thus recover the definition of the confined response function in its eigenbasis, as introduced in Sec. 2. Multiplying Eq.~\eqref{Dyson_gab} by $P$ on the left and on the right, we obtain two scalar equations for the symmetric and antisymmetric response functions: 
\beq
\left\{
\begin{array}{l}
g^{\rm s}_{\rm ew}(\tb{q},\omega)= -g_{\rm e}^{\rm s} \,g_{\rm w}^{\rm s}+g_{\rm e}^{\rm s}\, g_{\rm w}^{\rm s}\, g_{\rm ew}^{\rm s} \\
g^{\rm a}_{\rm ew}(\tb{q},\omega)= -g_{\rm a}^{\rm s} \,g_{\rm w}^{\rm a}+g_{\rm e}^{\rm a}\, g_{\rm w}^{\rm a}\, g_{\rm ew}^{\rm a} \\
\end{array}
\right.
\eeq
Once the Dyson equation is reduced to a scalar equation, we may follow the steps of refs. \cite{Kavokine2022,yu2023} to obtain the friction coefficient and thermal boundary conductance in the confined geometry: 
\beq\label{lambda_force}
\lambda =\frac{\hbarr^2}{8\pi^2T}\int_0^\infty\dd\omega\dd q\, \frac{q^3}{\sh^{2}\left(\frac{\hbarr\omega}{2T}\right)}( \Delta \gamma[F^{\rm s}] + \Delta \gamma[F^{\rm a}]),
\eeq
\beq
\kappa =\frac{\hbarr^2}{4\pi^2T^2}\int_0^\infty\dd\omega\dd q\, \frac{q\omega^2}{\sh^{2}\left(\frac{\hbarr\omega}{2T}\right)} ( \Delta \gamma[F^{\rm s}] + \Delta \gamma[F^{\rm a}]).
\eeq
The confined response functions thus emerge naturally in the theory of fluctuation-induced effects in a 2D channel. 

 \begin{figure*}
     \centering
     \includegraphics[scale=1]{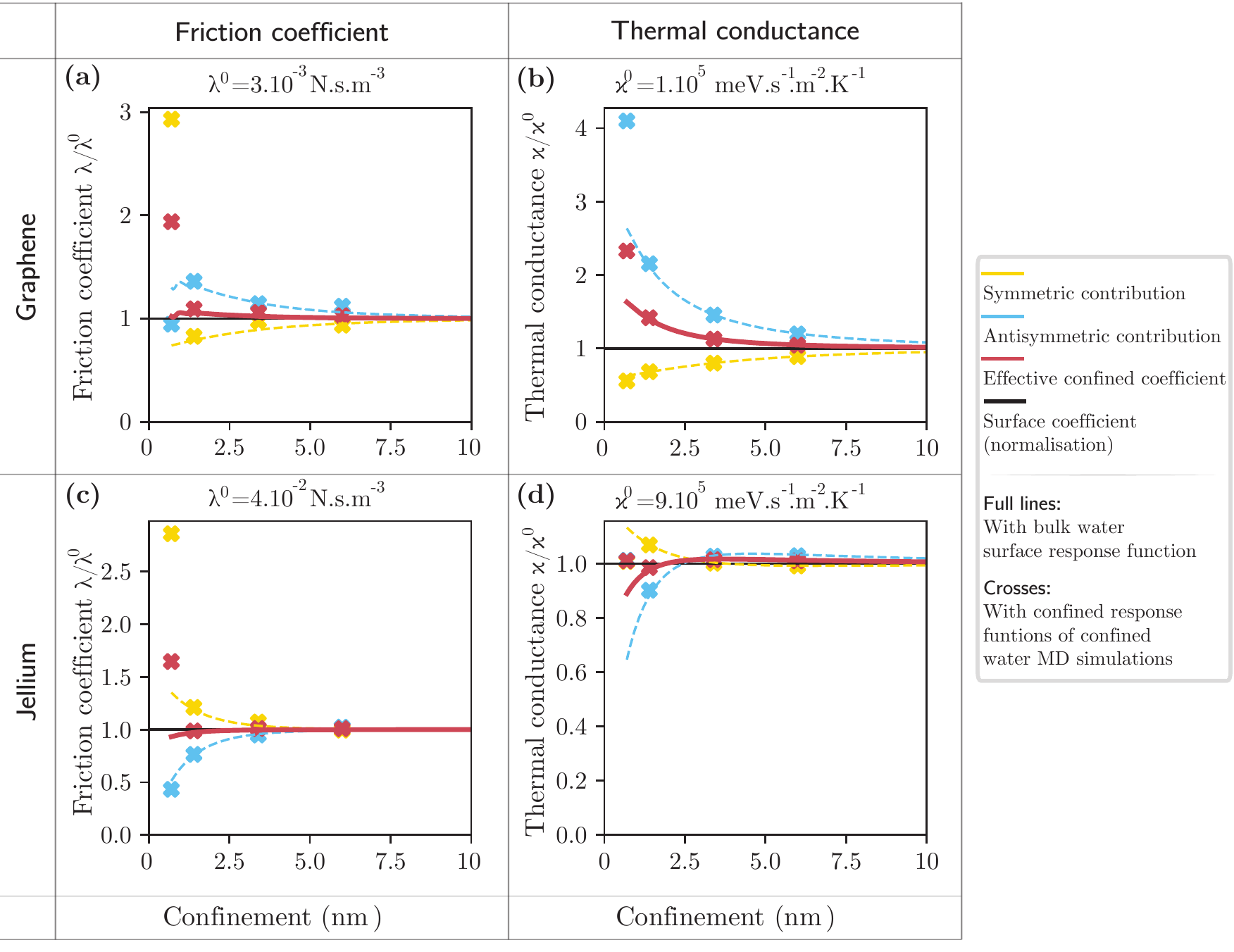}

     \caption{\emph{Effect of confinement on the fluctuation-induced phenomena.} a) Quantum friction coefficient $\lambda$ normalised by the single interface quantum friction coefficient $\lambda^0$ as a function of confinement between graphene walls. 
     b) Thermal boundary conductance $\kappa$ normalised by the single interface thermal boundary conductance $\kappa^0$ as a function of confinement between graphene walls. 
     c) Quantum friction coefficient $\lambda$ normalised by the single interface quantum friction coefficient $\lambda^0$ as a function of confinement between jellium walls with an effective mass $m~=~0.1~m_{\rm e}$ and a Fermi level $\mu~=~180$~meV (electron density parameter $r_s~=~5$)
          d) Thermal boundary conductance $\kappa$ normalised by the single interface thermal boundary conductance $\kappa^0$ as a function of confinement between jellium walls  with an effective mass $m~=~0.1~m_{\rm e}$ and a Fermi level $\mu~=~180$~meV (electron density parameter $r_s~=~5$).  In all panels, the continuous lines are obtained using the single interface surface response function for water, while the crosses are obtained with the confined response function of water at the relevant confinement, as obtained from molecular dynamics simulations.    } 
          \label{fig5}
 \end{figure*}

\subsection{Diagrammatic approach to confined response function}\label{diagconf}

Using the diagrammatic approach developed for the fluctuation-induced effects, we may interpret the confined response functions of the solid walls as surface response functions that have been renormalised by the inter-wall interactions within the random phase approximation (RPA). 

We start from the intra-wall response function $\chi_{\rm e}(z,z')$, that has been renormalized by the intra-wall Coulomb interactions at the RPA level, according to the Dyson equation shown diagrammatically in Fig. 4c. These are identical in the top and bottom walls ($\chi_{\rm e}^{\rm BB} = \chi_{\rm e}^{\rm TT}$), and, at this stage, there is no inter-wall response $\chi_{\rm e}^{\rm BT}$, since we do not allow for electron tunneling between the walls. 

In the presence of Coulomb interactions between the walls, we may introduce the exchange term $\Pi=\chi_{\rm e}^{\rm BB}\ast \chi_{\rm e}^{\rm TT}$ (Fig. 4e). Still at the RPA level, the intra-wall response function is then renormalised according to (Fig. 4f)
\beq \tilde{\chi}_{\rm e}^{\rm BB}=\chi_{\rm e}^{\rm BB}+\Pi\ast\tilde{\chi}_{\rm e}^{\rm BB}.\eeq 
The inter-wall response is no longer vanishing, and satisfies (Fig. 4g)
\beq\tilde{\chi}_{\rm e}^{\rm BT}=\Pi +\Pi\ast \tilde{\chi}^{\rm BT}_{\rm e}.\eeq 
Fourier-transforming these Dyson equations as detailed above, we obtain relations between confined and surface response functions. Using in particular that 
\beq
    \frac{ e^2}{2\epsilon_0 q}\int_{\mathfrak{Z}^{\rm B}} \dd z\int_{\mathfrak{Z}^{\rm T}}  \dd z'\, \Pi (z,z')e^{-q(z-z'-h)}= \left(g_{\rm e}^0\right)^2e^{-qh},
\eeq
we obtain 
\beq 
    g_{\rm e}^{\rm BB}=\frac{g_{\rm e}^0}{1-\left(g_{\rm e}^0\right)^2e^{-2qh}}, \qquad g_{\rm e}^{\rm BT}=-\frac{\left(g_{\rm e}^0\right)^2e^{-qh}}{1-\left(g_{\rm e}^0\right)^2e^{-2qh}}.
\eeq 
and deduce the symmetric and antisymmetric components of the confined response function
\beq\label{ge_confined}
g^{\rm s}_{\rm e} =
 \frac{g_{\rm e}^0}{1+g_{\rm e}^0e^{-qh}},
 \quad  \quad 
 g^{\rm a}_{\rm e}
= \frac{g_{\rm e}^0}{1-g_{\rm e}^0e^{-qh}}.
\eeq 
We thus recover Eq. \eqref{gco}, that we previously obtained from purely electrodynamic considerations.

\subsection{Effect of confinement on the fluctuation-induced effects}

We now investigate the effect of confinement on the quantum friction coefficient and thermal boundary conductance of water in a 2D channel, with walls made of either graphene or a semi-infinite jellium, with the parameters detailed in Sec. \ref{solid-water}. The results are presented in Fig. 5. 

We first focus on the effect of inter-solid interactions and thus evaluate the fluctuation-induced effects using the single-interface surface response function for water (continuous lines in Fig. 5). Interestingly, we observe opposite trends for graphene and for jellium walls. In the case of graphene, for both friction and thermal conductance, the antisymmetric contribution is enhanced and the symmetric contribution is reduced with confinement. Indeed, both effects are governed by the coupled plasmon modes of the walls, and the out-of-phase mode has lower energy than the in-phase mode, thus making a larger contribution. For our jellium model, the plasmon is well above the thermal energy (around 300 meV), and the fluctuation-induced effects are governed by single-particle excitations: we find that, in this case, the confinement enhances the symmetric contribution and reduces the antisymmetric contribution. The antisymmetric contribution dominates the behaviour of the total friction and thermal conductance, but the overall confinement effect remains lower than 10\%, except for the thermal conductance with graphene walls, where it reaches 50\%. In general, the confinement effect is stronger for graphene walls than for jellium walls because the electronic fluctuations that mediate quantum friction and near-field heat transfer have a longer wavelength (smaller momentum $q$) in the case of graphene. 

We now turn to the effect of the confinement-induced changes in the water fluctuations. Our simulations have shown that these changes become significant only at 7 \AA~confinement (Fig. 3), with an amplification of the Debye peak in the symmetric response and of the libration peak in the antisymmetric response. This translates into an enhancement of the friction coefficient and thermal conductance for both solid models, by up to a factor of 2 in the case of graphene. The study of fluctuation-induced effects specifically in 7 \AA~confinement is thus of particular interest. For instance, the thermal conductance of the interface between graphene and nanoconfined water may be probed with optical pump - terahertz probe spectroscopy: the optically-excited graphene electrons would be expected to cool faster than in the non-confined case~ \cite{yu2023}.

\section{Conclusions}\label{conclusion}

In this paper, we have developed a theoretical framework for studying confinement- and fluctuation-induced effects in two-dimensional nanofluidic channels: effects of (fluctuating) Coulomb interactions between the liquid and the solid. The key element of our framework is the description of the response of the solid walls and of the confined liquid to the Coulomb potentials that they apply to each other. In the case of a single solid-liquid interface, the surface response function -- the reflection coefficient for evanescent plane waves -- was found to be the most convenient descriptor. This convenience was due to the evanescent plane waves being eigenmodes of the Coulomb potential: the response to an evanescent wave is an evanescent wave. Generalizing this idea to the 2D channel geometry, we have introduced \emph{confined response functions}, that play the role of reflection coefficients for the potential eigenmodes of the confined system. Our approach is systematic, and potentially extendable to more complex geometries. 

The confined response functions reveal electrodynamic cross-talk between the walls of a 2D nanochannel. Investigating model materials that exhibit a surface plasmon mode, we found that the plasmons of the two walls couple as soon as the confinement is comparable to the plasmon wavelength. While the coupling of collective modes through Coulomb interactions is in principle a well-known phenomenon \cite{Hwang2010}, our framework allows for the investigation of its effect on nanoscale fluid transport. From the fluid side, we have investigated confined water through molecular dynamics simulations. We found that the water confined response remains essentially bulk-like in the thermal frequency range down to 1.4 nm confinement, but undergoes significant changes when the confinement reaches 7 \AA. 

As an application of our framework, we have investigated quantum friction and near-field radiative heat transfer between water and the walls of a 2D nanochannel. We have generalized the derivation of refs.~ \cite{Kavokine2022,yu2023} to a the confined geometry and found that the confined response functions naturally emerge. In channels wider than 7~\AA, the friction coefficient and thermal boundary conductance are modified compared to their bulk values when the confinement is comparable to the typical wavelength of the relevant charge fluctuations. At 7~\AA~confinement, a significant enhancement occurs for both effets, due to the drastic modification of the water response. Observing a confinement-induced modification of quantum friction or heat transfer thus appears most promising for systems where charge fluctuations are on longer wavelengths, yet these are also the systems where the effects are the weakest. Nevertheless, it has been shown that graphene-water heat transfer can be measured with ultrafast spectroscopy \cite{yu2023}, and a small quantum friction coefficient can still result in a large quantum friction force (termed quantum feedback) if the electrons are driven by a phonon wind \cite{Coquinot2023}. Our results thus provide guidelines for engineering fluctuation-induced effects in nanoscale fluid transport. 

\section*{Author Contributions}
N.K. and L.B. designed the project. B.C. carried out the theoretical analysis. M.B. performed the molecular dynamics simulations, with supervision from R.N. B.C. and N.K. co-wrote the paper, with input from all authors. All authors discussed the results and commented on the manuscript. 

\section*{Conflicts of interest}
There are no conflicts to declare.

\section*{Acknowledgements}
The Flatiron Institute is a division of the Simons Foundation.  B.C. acknowledges funding from a J.-P. Aguilar grant of the CFM Foundation, and thanks the Flatiron Institute and the Max Planck Institute for Polymer Research for their hospitality. M.B and R.R.N acknowledge funding by the Deutsche Forschungsgemeinschaft (DFG, German Research Foundation) - Project-ID 387284271 - SFB 1349. L.B. acknowledges funding from the EU H2020 Framework Programme/ERC Synergy Grant agreement number 101071937 n-AQUA. N.K. acknowledges support from a Humboldt fellowship. We thank Lucy Reading-Ikkanda (Simons Foundation) for help with figure preparation. 




\bibliography{bibfile} 
\bibliographystyle{rsc} 

\end{document}


\title{Collective modes and quantum effects in two-dimensional nanofluidic channels\\
\ti{Electronic Supplementary Information (ESI)}}

\author{Baptiste Coquinot,\textit{$^{a,b,c}$} Maximilian Becker,\textit{$^{d}$} Roland R. Netz,\textit{$^{d}$} Lyd\'eric Bocquet\textit{$^{a}$} and Nikita Kavokine\textit{$^{b,c}$}}
\email{Contact: nikita.kavokine@mpip-mainz.mpg.de.}

\affiliation{ $^{a}$~Laboratoire de Physique de l'\'Ecole Normale Sup\'erieure, ENS, Universit\'e PSL, CNRS, Sorbonne Universit\'e, Universit\'e Paris Cit\'e, 24 rue Lhomond, 75005 Paris, France.\\
$^{b}$~Department of Molecular Spectroscopy, Max Planck Institute for Polymer Research, Ackermannweg 10, 55128 Mainz, Germany.\\
$^{c}$~Center for Computational Quantum Physics,
Flatiron Institute, 162 5$^{th}$ Avenue, New York, NY 10010, USA.\\
$^{d}$~Fachbereich Physik, Freie Universitat Berlin, Arnimallee 14, 14195 Berlin, Germany.}


\maketitle
\tableofcontents


\section{Absence of overscreening with the confined response functions}\label{SI_solid_RF_2}

\label{overscreening}

Since the real part of the surface response functions consists of positive numbers smaller than one, Eq.~\ts{(16)} of the main text allows the antisymmetric response function to become larger than 1. This is in particular the case at zero frequency when the imaginary parts vanish. 

Physically, it is easy to understand that the symmetric response function should be smaller than the surface response function while its antisymmetric counterpart should be larger. Indeed, let us focus on the top solid, which responds to the sum of the (positive) external potential and the potential induced by the bottom solid. In the symmetric case the bottom solid also responds to a positive external potential and then induces a negative potential which is damped exponentially. However, the surviving part reaching the top solid is still negative and then reduces the effective external potential to which the top solid responds. The symmetric response is then smaller than the surface response because the external potential is screened by the bottom solid. On the contrary, in the antisymmetric case, the external potential is negative on the bottom solid which then responds by inducing a positive field. By the same process, the top solid now sees an effective external potential which is larger, generating a larger response. The antisymmetric response should then be larger than the surface response.

However, this cannot generate overscreening. Indeed,  even if the antisymmetric response function is larger than 1, one should remember that the induced potential is not of the form of the external potential but of its "conjugate". The antisymmetric inner weight function is smaller than its outer counterpart: this solves the apparent problem. Let us formalise this intuition by computing the ratio of the induced field by the external field between the interfaces:
\beq -\frac{ \phi_{\rm ind}\left(z\right)}{ \phi_{\rm ext}\left(z\right)}=g_{\rm e}^{\rm x}\frac{ F_{\rm i}^{\rm x}(z)}{F_{\rm o}^{\rm x}(z)}=\frac{g_{\rm o}e^{-qh}}{1+\eta g_{\rm o}e^{-qh}}\frac{|e^{qz}+\eta e^{-qz}|}{e^{-q|z|}}.\eeq
where $\eta=\pm1$ depending on whether we consider the symmetric or antisymmetric case. This ratio is maximal for $z=\pm h/2$ where it reaches
\beq -\frac{ \phi_{\rm ind}\left(\pm h/2\right)}{ \phi_{\rm ext}\left(\pm h/2\right)}=g_{\rm o}\frac{1+\eta e^{-qh}}{1+\eta g_{\rm o}e^{-qh}}=1-\frac{1-g_{\rm o}}{1+\eta g_{\rm o}e^{-qh}}\leq 1. \eeq
Therefore, the induced potential is always smaller than the external potential, even when the confined response function is larger than 1. There is no overscreening. 

As a consequence, the water static confined response function must fulfill the inequality $g_{\rm w}^{\rm x}\leq 1+ \eta e^{-qh}$. In particular, the static antisymmetric response function goes to zero at $q\rightarrow 0$. For comparison, its surface response function has the limit $g_{\rm w}(\omega=0, q\rightarrow 0)\rightarrow (\epsilon-1)/(\epsilon+1)\approx 0.98$ (see \cite{Kavokine2022}).

\section{Electrostatic interaction confinement}\label{interaction_confinement}

\begin{figure*}
    \centering
	\includegraphics{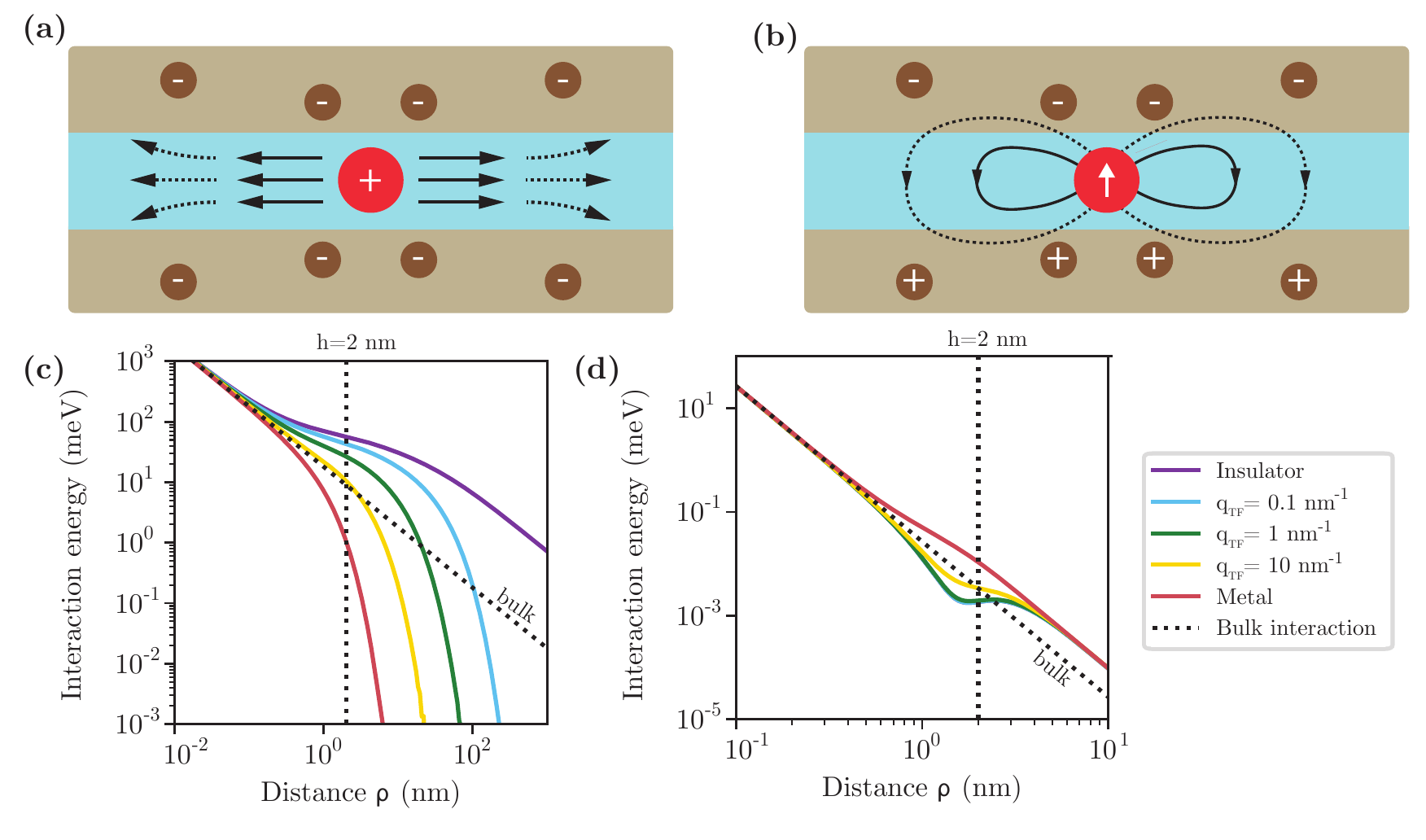}
	
    \caption{ a) Interaction confinement of an ion in the center of the channel: the generated electric field is partly confined in the channel by the surrounding solid which develop counter-charges (pictured in dark brown). This leads to a modified ion-ion interactions. b) Interaction confinement of an electrostatic dipole in the center of the channel: the generated electric field is partly confined in the channel by the surrounding solid which develop counter-charges (pictured in dark brown). This leads to a modified dipole-dipole interactions. c) Ion-ion interaction energy shift (normalised by the bulk interaction energy) for two ions at a distance $\rho$ in bulk (black dashed line) and in a channel of confinement $h=2$ nm. Different models of solid are used. d) Dipole-dipole interaction energy shift (normalised by the bulk interaction energy) for two dipoles in the $z$-direction  at a distance $\rho$ in bulk (black dashed line) and in a channel of confinement $h=2$ nm. Different models of solid are used.}
    \label{fig1bis}
\end{figure*}

  \begin{table*}
 \centering
 \renewcommand{\arraystretch}{2}
\begin{tabular}{ l @{\hspace{0.5cm}} l @{\hspace{0.5cm}} @{\hspace{0.5cm}}c @{\hspace{0.5cm}} @{\hspace{0.5cm}}c@{\hspace{0.5cm}} }
    \hline  
    \tb{Model} & \tb{Geometry} & \multicolumn{2}{c}{\tb{Weight function}}  \\
    \hline  
    \tb{Single interface} & \underline{Half-space}  & \multicolumn{2}{c}{$F^0(q,z)=e^{-q|z|}$}   \\ \hline 
   \tb{Confined} &  & \underline{Symmetric} &   \underline{Antisymmetric}  \\
     & \underline{Inside} &$F^{\rm s}_{\rm i}(q,z)=\sqrt{2}\,\ch(qz)e^{-qh/2}$ & $F^{\rm a}_{\rm i}(q,z)=\sqrt{2}\,\sh(qz)e^{-qh/2}$\\
     & \underline{Outside} &  $F^{\rm s}_{\rm o}(q,z)=\frac{1}{\sqrt{2}}e^{-q(|z|-h/2)}$ & $F^{\rm a}_{\rm o}(q,z)=\frac{\tn{sign}(z)}{\sqrt{2}}e^{-q(|z|-h/2)}$\\
     \hline 
     \end{tabular}
\caption{Weight functions used in the definitions of the surface and confined response functions (Eq. \eqref{def_g0} and \eqref{IRF}).}\label{able}
\end{table*}

\subsection{Model and solid's static response function }

In this section, we focus on how confinement will modify the electrostatic interactions between charges inside a channel. While our formalism allows us to treat the electric interactions of any static or dynamic distribution of charge, we restrict here to basic cases. We consider either an ion or a dipole in the center of the channel, as represented in Fig.~\ref{fig1bis}a-b, to deduce the ion-ion and the dipole-dipole electrostatic interactions in presence of confinement.

To apply our theory and to compare with the existing literature, we need a simple model of the solid' surface response function. For this, we use the work of Kavokine \ti{et al.} \cite{Kavokine2022b}. To start, let us duplicate their eye-opening derivation of the surface response function of an insulator of static dielectric constant $\epsilon_{\rm m}\approx 2$, taking into account the presence of water of static dielectric constant $\epsilon_{\rm w}\approx 80$ at the interface.  

We consider an external electric potential $\phi_{\rm ext}(\q,z)=\phi_{\rm ext}^0(\q)e^{qz}$ applied on the solid in $z<0$. On the solid area the total electric potential is denoted $\phi_{\rm m}(\q,z)=\phi_{\rm m}^0(\q)e^{qz}$ and the solid generates an induced electric potential  $\phi_{\rm ind}(\q,z)=\phi_{\rm m}^0(\q)e^{-qz}$ on the area $z>0$. The boundary conditions at $z=0$ are the continuity of both the total electric potential $\phi$ and the electric displacement field $\tb{D}=-\epsilon\nabla\phi$, leading to
\beqa \phi_{\rm ext}^0+\phi_{\rm ind}^0&=&\phi_{\rm m}^0\\
 \epsilon_{\rm w}\left(\phi_{\rm ext}^0-\phi_{\rm ind}^0\right)&=&\epsilon_{\rm m}\phi_{\rm m}^0.\eeqa
 Finally, we obtain the surface response function:
 \beq
 g_{\rm o}^0(\q)=\frac{\phi_{\rm ind}^0(\q)}{\phi_{\rm ext}^0(\q)}=\frac{ \epsilon_{\rm m}- \epsilon_{\rm w}}{ \epsilon_{\rm m}+ \epsilon_{\rm w}}\eeq
 This derivation is valid for an insulator, while the dielectric constant of the solid depends on the momentum $q$ in general.  A generalisation of this formula exists for a Thomas-Fermi model of solid, parametrised with a Thomas-Fermi wavelength $q_{\rm TF}$ \cite{Kavokine2022b}:
  \beq \label{g_static}
 g_{\rm o}^0(\q)=\frac{ \epsilon_{\rm m}f_{\rm TF}(q)- \epsilon_{\rm w}}{ \epsilon_{\rm m}f_{\rm TF}(q)+ \epsilon_{\rm w}}, \quad f_{\rm TF}(q)=\sqrt{1+\frac{1}{\epsilon_{\rm m}}\left(\frac{q_{\rm TF}}{q}\right)^2}.\eeq
 The case $q_{\rm TF}=0$ corresponds to an insulator. The case $q_{\rm TF}\rightarrow \infty$, that is $g_{\rm o}^0(\q)\rightarrow 1$, corresponds to a metal. Let us highlight that $ g_{\rm m}^0$ is negative at large wavelength and positive at small wavelength. The wavelength of the transition depends on the Thomas-Fermi wavelength. When the surface response function is negative, which is expected to happen at small distances, the effect of the boundary solids is to confine the electric field inside the channel:  this is \emph{interaction confinement}. At larger distances, the surface response function becomes ultimately positive: the electric field is no longer confined.

\subsection{Interaction confinement for an ion and ion-ion electrostatic interaction}

Let us start with the case of a static ion of charge $e$ (see Fig.~\ref{fig1bis}a). Such a charge generates a potential 
\beq  \phi_{\rm ext}(\rho,z)=\int \frac{\dd\q}{(2\pi)^2}\frac{e^2}{2\epsilon q}\sqrt{2}F^{\rm s}_{\rm o}(q,z)e^{i\rho\cdot\q-qh/2}  \eeq 
where we have introduced a Fourier decomposition into independent modes. Taking into account the effect of the material, the total potential in the inner space now writes 
\beq  \phi(\q,z)=\frac{\sqrt{2}e^2}{2\epsilon q}\left[F^{\rm s}_{\rm o}(q,z)-g^{\rm s}_{\rm o}(\tb{q})F^{\rm s}_{\rm i}(q,z)\right]e^{-qh/2}  \eeq
In particular, in the center of the channel, that is at $z=0$, we obtain using Eq. \ts{(17)} of the main text: 
\beq  \phi(\q,z=0)=\frac{e^2}{2\epsilon q}\left[1-\frac{2g^0_{\rm o}(\q)e^{-qh}}{1+g^0_{\rm o}(\q)e^{-qh}}\right]  \eeq
consistently with the result of Kavokine \ti{et al.} \cite{Kavokine2022b}. An ion in the center of the channel is a symmetric distribution of charge and then its potential is corrected by the symmetric response function of the outer medium. The effect of the medium is to screen the charge and then to confine the electric potential, as pictured in Fig.~\ref{fig1bis}a.

Considering another ion of charge $e$ in the center of the channel at a distance $\rho$, the ion-ion electrostatic interaction is 
\beq \mc{E}_{\rm ii}=\phi(\rho,z=0)=\frac{e^2}{2\epsilon }\int_0^\infty \frac{\dd q}{2\pi}J_0(q\rho)\left[1-\frac{2g^0_{\rm o}(q)e^{-qh}}{1+g^0_{\rm o}(q)e^{-qh}}\right]
\eeq
where we have used isotropy and introduced the Bessel function of the first kind $J_0$. The first term provides te bulk interaction energy 
\beq \mc{E}_{\rm ii}^{\rm bulk}(\rho)=\frac{e^2}{4\pi\epsilon_0\epsilon_{\rm w} \rho}\eeq
while the second term provides a correction due to confinement. 
This correction can be computed numerically using Eq. \eqref{g_static}.
 The resulting ion-ion interaction (normalised by $\mc{E}_{\rm ii}^{\rm bulk}(\rho)$) is plotted in Fig.~\ref{fig1bis}c. Consistently with \cite{Kavokine2022b}, we find that the interaction is reduced at large distance and increased at small distance for a generic solid. 

\subsection{Interaction confinement for a dipole and dipole-dipole electrostatic interaction}

Let us now turn to an electrostatic dipole $\tb{d}$ oriented in the $z$ direction (see Fig.~\ref{fig1bis}b). Such a dipole generates a potential 
\beq  \phi_{\rm ext}(\rho,z)=-\int \frac{\dd\q}{(2\pi)^2}\frac{e^2d}{2\epsilon }\sqrt{2}F^{\rm a}_{\rm o}(q,z)e^{i\rho\cdot\q-qh/2}  \eeq 
where we have introduced a Fourier decomposition into independent modes. Taking into account the effect of the material, the total potential in the inner space now writes 
\beq  \phi(\q,z)=-\frac{\sqrt{2}e^2d}{2\epsilon }\left[F^{\rm a}_{\rm o}(q,z)-g^{\rm a}_{\rm o}(\tb{q})F^{\rm a}_{\rm i}(q,z)\right]e^{-qh/2}  \eeq
In particular, in the center of the channel, using Eq. \ts{(17)} of the main text, the electric field writes
\beq  \tb{E}(\q,z=0)=-\frac{e^2d}{2\epsilon }q\tb{d}\left[1+\frac{2g^0_{\rm o}(\q)e^{-qh}}{1-g^0_{\rm o}(\q)e^{-qh}}\right]  \eeq
An electrostatic dipole in the center of the channel is an antisymmetric distribution of charge and then its potential is corrected by the antisymmetric response function of the outer medium. Here again, the effect of the outer medium is to confine the electric potential, as pictured in Fig.~\ref{fig1bis}b.

Considering another dipole of same moment $\tb{d}$ in the center of the channel at a distance $\rho$, the dipole-dipole electrostatic interaction is 
\begin{widetext}
	\beq \mc{E}_{\rm dd}=-\tb{d}\cdot\tb{E}(\rho,z=0)=\frac{e^2d^2}{2\epsilon }\int_0^\infty \frac{\dd q}{2\pi}q^2J_0(q\rho)\left[1+\frac{2g^0_{\rm o}(q)e^{-qh}}{1-g^0_{\rm o}(q)e^{-qh}}\right]
\eeq
\end{widetext}
The first term provides te bulk interaction energy 
\beq \mc{E}_{\rm dd}^{\rm bulk}(\rho)=\frac{e^2d^2}{4\pi\epsilon_0\epsilon_{\rm w} \rho^3}\eeq
while the second term provides a correction due to confinement. 
This correction can be computed numerically using Eq. \eqref{g_static}.
 The resulting dipole-dipole interaction (normalised by $\mc{E}_{\rm dd}^{\rm bulk}(\rho)$) is plotted in Fig.~\ref{fig1bis}d. After a reduction of the interaction at small distances compared with the confinement $h$, the interaction energy is increased by a factor $\approx 2.6$ independent on the model of solid.  

To conclude this section, we have derived the correction of confinement to the electric potential. This approach generalises the results of interaction confinement \cite{Kavokine2022b} to the generic distribution of charge, or equivalently to a generic external potential, which we can decompose into a symmetric and an antisymmetric parts.

\section{Water response functions and simulations}

\subsection{Numerical methods for water simulations}\label{numerical}

We have carried out both force-field (FF) and \ti{ab initio} density functional theory-based (DFT) molecular dynamics (MD) simulations of water
confined between two frozen graphene sheets for various separations $h$ between the graphene sheets. 

FF-MD simulations are performed using GROMACS 2020 using the SPC/E water model and GROMOS 53A6 parameters for carbon atoms \cite{Abraham2015, Oostenbrink2004}. Simulations are performed in the NVT ensemble using the CSVR thermostat at 300~K. We use a 2~fs time step and truncate Lennard-Jones interactions at 9~\AA. 
Electrostatics employ a real-space cut-off at 0.9~nm while long-range interactions are handled with the particle mesh Ewald method. 
The considered systems have lateral dimensions of 42.6~\AA~$\times$ 44.3~\AA~and contain 210, 1872 and 3489 water molecules at  $h=$7~\AA,  34~\AA~and 60~\AA graphene sheet separations, respectively. 
After 1 ns of equilibration, we perform production runs of 20~ns. 

DFT-MD simulations are carried out with the CP2K 6.1 software which performs Born-Oppenheimer MD \cite{Kuhne2020}. 
Electronic structures are calculated at every time step based on the BLYP exchange correlation functional with Grimme-D3 dispersion correction \cite{Grimme2010},
 while core electrons are represented by GTH pseudo potentials and valence electrons are expanded in the DZVP-SR-MOLOPT basis set using a plane wave cutoff of 400 Ry  \cite{Goedecker1996,VandeVondele2007}. DFT-MD simulations are performed in the NVT ensemble at 300K with a 0.5~fs time step using the CSVR thermostat. 
 The simulated systems have lateral dimensions of 29.82~\AA~$\times$~27.05~\AA~and contain 90 and 420 water molecules for graphene sheet separations
  of $h=$7 and 17.8~\AA, respectively. 
  After preequilibration in FF-MD, DFT-simulations are equilibrated for another 5~ps followed  by production runs of 70~ps.

\subsection{Fluctuation-dissipation theorem } \label{SI_gh}

Our goal is to compute the surface response function defined in Eq.~\ts{(8)} of the main text as
\beq
g(\tb{q},\omega)=-\frac{e^2}{2\epsilon_0 q}\int\dd z\dd z'\, F(q,z)\chi(z,z',\tb{q},\omega)F(q,z').
\eeq
Using the fluctuation-dissipation theorem we then deduce 
\beq \label{FD}
\im{g_{\rm w}[F](\tb{q},\omega)}=\frac{e^2}{4\epsilon_0 q}\frac{\omega}{T}S[F](\tb{q},\omega)\eeq
where 
\beq S[F](\tb{q},\omega)=\int_{-\infty}^\infty \langle\delta n_{\rm w}[F](\tb{q},t)\delta n_{\rm w}[F](-\tb{q},t)\rangle e^{i\omega t}\dd t\eeq
is a modified structure factor with en effective density
\beq n_{\rm w}[F](\tb{q},t)=\int \dd\tb{r}\, n_{\rm w}(\tb{r},t)F(q,z)e^{-i\tb{q}\cdot\tb{r}}. \eeq
The density $n_{\rm w}(\tb{r},t)$ is given by the simulation trajectory. 

\subsection{Evaluation of the time correlation function} \label{SI_gh}

For each time step, we first compute the Fourier-transformed charge density. From MD simulation, we have access to the center of each atom and use a partial charge $Z=\delta=0.41$e for the hydrogen atoms and $Z=-2\delta$ for the oxygen atoms. The density is then
\beq n(\tb{q})=\sum_j Z_j F(q, z_j)e^{iq_xx+iq_yy} \eeq
where $F$ is the weight function and $q_x, q_y$ are multiples of $2\pi/L_x$ ($2\pi/L_y$ respectively). 
We make this computation for $N_t=10^5$ time steps, and then obtain the fluctuating part of the density by subtracting the average
\beq \delta n(\tb{q},t)=n(\tb{q},t)-\langle n(\tb{q},t)\rangle_t\eeq 
where $\langle A(t)\rangle_t=\frac{1}{N_t}\sum_t A(t)$. 

We then carry out a fast Fourier transform (FFT): 
\beq \delta n(\tb{q},\omega)=\tn{FFT}[ n(\tb{q},t)]_t\eeq and only keep the positive frequencies.
The response function is then computed as
\beq \im{g(\tb{q},\omega)}= \frac{e^2\omega}{4\epsilon_0qT\mc{A}} |\delta n(\tb{q},\omega)|^2\frac{\dd t}{N_t}.
\eeq
where $\dd t$ is the length of the time step. In practice, $g$ only depends on the norm $q$.  Finally, the spectra are smoothed by a Gaussian smoothing of width 0.2~meV for the FF simulations and 0.5~meV for DFT simulations.

\subsection{Momentum dependence and position of the interface}\label{numerical}

The limited size of the simulation box does not allow us to access all momenta necessary for the computation of fluctuation-induced effects (typically below 1~\AA$^{-1}$). Therefore, we restrict ourselves to one momentum compatible with the size of the box and extrapolate the response functions to lower momenta. We use $q_0=0.67$~\AA$^{-1}$ for FF simulations and $q_0=1$~\AA$^{-1}$ for DFT simulations. 

From the simulations carried out with a large box in \cite{Kavokine2022}, we know that the non-confined water surface response function has only a weak momentum dependence: we therefore approximate the surface response function at all momenta by the one computed at $q_0$.  For the confined response functions, however, the form of the weight functions imply that the static ($\omega= 0$) response fulfills $g_{\rm w}^{\rm x}\leq 1+ \eta e^{-qh}$, where $\eta=\pm 1$ for the symmetric (antisymmetric) component (see. \ts{ESI.1}). In particular, the static antisymmetric response function goes to zero at $q\rightarrow 0$. We thus infer the momentum dependence of the confined response function according to 
\beq
g_{\rm w }^{\rm x} (q, \omega) = \frac{1+ \eta e^{-qh}}{1+ \eta e^{-q_0h}} g_{\rm w}^{\rm x} (q_0,\omega). 
\eeq
In practice, in the evaluation of friction coefficients, this procedure makes only a 1\% difference compared to the assumption of no momentum dependence. Formally, however, such a procedure allows us to obtain well-behaved response functions that produce no unphysical overscreening. 

The evaluation of the response functions requires to precisely define the positions of the interfaces at $z = \pm h/2$. In ref. \cite{Kavokine2022}, a procedure for positioning the interface based on imposing the compressibility sum rule for the surface response function was proposed. This procedure no longer formally holds in a confined geometry, but we expect the position of the interface to not be significantly affected by confinement. With our simulations at weakest confinement (FF at 60 \AA~and DFT at 34 \AA) and following the procedure of ref. \cite{Kavokine2022}, we an effective distance between the plane of the carbon atoms and the water surface $d\approx 1.4$~\AA, which is compatible with $d = 1.3$~\AA found in \cite{Kavokine2022}. We then adopt the prescription $d= 1.4$~\AA~for the simulations under confinement.

\bibliographystyle{naturemag}
\bibliography{bibfile}